\documentclass[twocolumn,10pt,showpacs,preprintnumbers,amsmath,amssymb]{revtex4}

\usepackage{graphicx}
\usepackage{graphics}
\usepackage{dcolumn}
\usepackage{bm}

\begin{document}

\title{Relaxation channels of two-vibron bound states in $\alpha$-helix proteins}

\author{V. Pouthier and C. Falvo}

\address{Laboratoire de Physique Mol\'{e}culaire, UMR CNRS 6624. Facult\'{e} des
Sciences - La Bouloie, \\ Universit\'{e} de Franche-Comt\'{e}, 25030 Besan\c {c}on cedex, 
France.}

\email{vincent.pouthier@univ-comte.fr}

\date{\today}

\begin{abstract}
Relaxation channels for two-vibron bound states in an anharmonic $\alpha$-helix protein are studied. According to a recently established small polaron model [V. Pouthier, Phys. Rev. \textbf{E68}, 021909 (2003)], it is pointed out that the relaxation originates in the interaction between the dressed anharmonic vibrons and the remaining phonons. This interaction is responsible for the occurrence of transitions between two-vibron eigenstates mediated by both phonon absorption and phonon emission. 
At biological temperature, it is shown that the relaxation rate does not significantly depends on the nature of the two-vibron state involved in the process. Therefore, the lifetime for both bound and free states is of the same order of magnitude and ranges between 0.1 and 1.0 ps for realistic parameters. By contrast, the relaxation channels strongly depend on the nature of the two-vibron states which is a consequence of the breather-like behavior of the two-vibron bound states. 
\end{abstract}

\pacs{03.65.Ge, 63.20.Ry, 63.22.+m, 87.15.-v}

\maketitle

\section{Introduction}

Since the pioneer works of Davydov and co-workers \cite{kn:davydov}, soliton mechanisms for bioenergy transport in proteins have received increasing attention during the last twenty five years \cite{kn:scott1,kn:chris}. 
The main idea is that the energy released by the hydrolysis of adenosine triphosphate (ATP) can be stored in the C=O vibration (amide-I) of a peptide group. The dipole-dipole coupling between peptide groups leads to the delocalization of these vibrations and to the formation of vibrational excitons, i.e. vibrons. Therefore, the strong interaction between the vibrons and the phonons of the protein yields a nonlinear dynamics which favors the occurrence of the so called Davydov's soliton. 

However, it has been pointed out that the solution of the Davydov's problem is rather a small vibron-polaron than a vibron-soliton \cite{kn:brown1,kn:brown2,kn:ivic1,kn:ivic2,kn:ivic3,kn:ivic4}. Indeed, the vibron bandwidth in proteins is smaller than the phonon cutoff frequency so that the non-adiabatic limit is reached. During its propagation, a vibron is dressed by a virtual cloud of phonons which yields a lattice distortion essentially located on a single site and which follows instantaneously the vibron (small polaron). Nevertheless, the dressing effect leads to an attractive interaction between vibrons mediated by virtual phonons. Such an interaction is responsible for the formation of bound states and it has been suggested that proteins can support solitons formed by bound states involving a large number of vibrational quanta \cite{kn:ivic2,kn:ivic3,kn:ivic4}. 

Although this formalism gives a comprehensive schema for the formation of solitons in proteins, it assumes the harmonic approximation for the amide-I vibration. However, this approximation failed when several vibrons are excited because the intramolecular anharmonicity acts as an additional nonlinear source.
As the dressing effect, the anharmonicity is responsible for the formation of bound states \cite{kn:kimball,kn:bogani,kn:scott,kn:pouthier1,kn:pouthier2,kn:pouthier3} and the fundamental question of the interplay between both sources of nonlinearity has been addressed in a recent paper \cite{kn:pouthier4}. In this work, we have restricted our attention to the formation of two-vibron bound states (TVBS's), only. Indeed, although the influence of the anharmonicity in molecular lattices have been the subject of intense research during the last decade, this research was essentially restricted to classical lattices (for a recent review see for instance Refs. \cite{kn:aubry,kn:flach,kn:mackay}). In particular, the formation of discrete breathers, i.e. highly localized nonlinear vibrations, has been demonstrated. However, in spite of the great interest that these classical nonlinear objects have attracted, no clear evidence has been found for their existence in real lattices. By contrast, TVBS are have been observed in several low-dimensional molecular lattices \cite{kn:guyot1,kn:sih,kn:shen,kn:jakob1,kn:jakob2,kn:jakob3,kn:jakob4,kn:jakob_p1,kn:okuyama}. These quantum objects correspond to the first quantum states which experience the nonlinearity and can thus be viewed as the quantum counterpart of breathers or soliton excitations \cite{kn:scott}. Their characterization is thus essential and appears as a first step to understand the formation of multi-vibron solitons.

In Ref.\cite{kn:pouthier4}, we have shown that the anharmonicity modifies the vibron-phonon interaction which results in an enhancement of the dressing effect. Anharmonic vibrons are thus more sensitive to the dressing than harmonic vibrons. Moreover, both nonlinear sources favor the occurrence of two kinds of bound states which the properties strongly depend on the anharmonicity. In the harmonic situation, the two bound states appear as combinations of states involving the trapping of the two vibrons onto the same amide-I mode and onto nearest neighbor amide-I modes. By contrast, the intramolecular anharmonicity reduces the hybridization between these two kinds of trapping so that the low frequency bound state refers to the trapping of the two vibrons onto the same amide-I mode whereas the high frequency bound state characterizes the trapping onto nearest neighbor amide-I vibrations. 

In this study, the dynamical coupling between the dressed anharmonic vibrons and the remaining phonons was disregarded. Therefore, the present paper is devoted to the characterization of this coupling and to a detailed analysis of the relaxation pathways. 
The TVBS lifetime is determined with a special emphasis on the influence of the different nonlinear sources. 

At biological temperature, the lifetime of the Davydov's soliton is still an open question. It has been shown that the amide-I excitation, in vivo, corresponds to a localized state \cite{kn:cruzeiro1,kn:cruzeiro2}. Instead of traveling in a coherent manner, it follows a stochastic, diffusional path along the lattice. In other words, the single-vibron Davydov soliton does not last long enough to be useful at biological temperatures and it has been shown that two-vibron solitons are more stable and appear as good candidates for bioenergy transport \cite{kn:cruzeiro3,kn:forner}. However, recent calculations performed by Ivic et al. \cite{kn:ivic4} clearly show that the multi-vibron soliton lifetime is of about a few picoseconds, i.e. the same order of magnitude as the single-vibron soliton lifetime found by Cottingham and Schweitzer \cite{kn:cottingham,kn:schweitzer}.

The paper is organized as follows. In Sec. II, the dressed anharmonic vibron point of view described in details in Ref. \cite{kn:pouthier4} is first summarized. Then, the coupling Hamiltonian between these anharmonic polarons and the remaining phonons is determined. The TVBS relaxation rate is expressed in Sec. III and studied numerically in Sec. IV. The results are finally discussed in Sec. V.    

\section{Vibron-phonon Hamiltonians and two-vibron eigenstates}
 
\subsection{The general vibron-phonon Hamiltonian}

According to the original Davydov's model, the collective dynamics of the amide-I vibrations is described by a one-dimensional lattice with $N$ sites containing the C=O vibrations. The $n$th amide-I mode is assumed to behave as a high frequency anharmonic oscillator described by the standard operators $b^{+}_{n}$ and $b_{n}$. This oscillator is characterized by its harmonic frequency $\omega_{0}$ and by the cubic and quartic anharmonic parameters $\gamma_{3}$ and $\gamma_{4}$, respectively. Finally, the dipole-dipole coupling between nearest neighbor amide-I modes is described by the constant $J$. These C=O vibrations interact with the phonons of the lattice which characterize the dynamics of the external motions of the peptide groups. Within the harmonic approximation, the phonons correspond to a set of $N$ low frequency acoustic modes labeled $\{ q \}$ and described by the phonon operators $a^{+}_{q}$ and $a_{q}$. The frequency of the $q$th mode is defined as $\Omega_{q}=\Omega_{c}\mid sin(q/2) \mid$, where $\Omega_{c}$ denotes the phonon cutoff frequency.

As shown in Ref. \cite{kn:pouthier4}, a unitary transformation is performed to remove the intramolecular anharmonicity of each amide-I mode and a modified Lang-Firsov transformation is applied to renormalize the vibron-phonon interaction. As a result, the vibron-phonon Hamiltonian is defined as (within the convention $\hbar=1$)
\begin{eqnarray}
&&\hat{H}=\sum_{n} \hat{\omega}_{0}b^{+}_{n}b_{n}-\hat{A}b^{+2}_{n}b_{n}^{2}-\hat{B}b^{+}_{n+1}b^{+}_{n}b_{n+1}b_{n}  \nonumber \\
&-&J_{1}[\Theta_{n}^{+}(N_{n}-1)\Theta_{n+1}(N_{n+1})b^{+}_{n}b_{n+1}+H.c.] \nonumber \\
&-&J_{2}[\Theta_{n}^{+2}(N_{n}-\frac{3}{2})\Theta_{n+1}^{2}(N_{n+1}+\frac{1}{2})b^{+2}_{n}b_{n+1}^{2}+H.c.] \nonumber \\
&-&J_{3}[\Theta_{n}^{+}(N_{n}-1)\Theta_{n+1}(N_{n+1})b^{+}_{n}[N_{n}+N_{n+1}]b_{n+1}+H.c.] \nonumber \\
&+&\sum_{q} \Omega_{q}a^{+}_{q}a_{q} 
\label{eq:HHAT}
\end{eqnarray}
where $N_{n}=b_{n}^{+}b_{n}$ and $A=30\gamma_{3}^{2}/\omega_{0}-6 \gamma_{4}$. In Eq.(\ref{eq:HHAT}), H.c. stands for the Hermitian conjugate and the different parameters  are expressed in terms of both the anharmonic parameters and the small polaron binding energy $E_{B}$ as
\begin{eqnarray}
\hat{\omega}_{0}&=&\omega_{0}-2A-B-(1+4\eta)E_{B} \nonumber \\
\hat{A}&=&A+(1+8\eta)E_{B} ; \hat{B}=B+(1+4\eta)E_{B} \nonumber \\
B&=&144 J (\frac{\gamma_{3}}{\omega_{0}})^{2} 
; J_{1}=J(1+44(\frac{\gamma_{3}}{\omega_{0}})^{2}-12\frac{\gamma_{4}}{\omega_{0}}) \nonumber \\
J_{2}&=&4 J(\frac{\gamma_{3}}{\omega_{0}})^{2} 
; J_{3}=J(22(\frac{\gamma_{3}}{\omega_{0}})^{2}-12\frac{\gamma_{4}}{\omega_{0}}) \nonumber \\
\eta&=&120(\frac{\gamma_{3}}{\omega_{0}})^{2}-12\frac{\gamma_{4}}{\omega_{0}} 
\label{eq:parameter2}
\end{eqnarray}
Note that the small polaron binding energy $E_{B}$ and the anharmonic parameter $A$ appear as the relevant parameters to characterize the nonlinearity of the system.
In Eq.(\ref{eq:HHAT}), $\Theta_{n}(N_{n})$ stands for the dressing operator expressed as
\begin{equation}
\Theta_{n}(N_{n})=exp(-Q_{n}[1+2\eta+2\eta N_{n}])
\label{eq:theta}
\end{equation}
where $Q_{n}$ is defined as
\begin{equation}
Q_{n}=\sum_{q} \sqrt{\frac{E_{B}}{2N\Omega_{q}}}\frac{sin(q)}{i\mid sin(q/2) \mid} e^{-iqn}a_{q}^{+}-H.c
\label{eq:Q}
\end{equation}

The Hamiltonian Eq.(\ref{eq:HHAT}) describes the dynamics of anharmonic vibrons dressed by virtual phonons, i.e. anharmonic small polarons. It takes into account on the intramolecular anharmonicity up to the second order and allows for a renormalization of the main part of the vibron-phonon coupling. However, this coupling remains through the dressing operators $\Theta_{n}(N_{n})$ which depend on the phonon coordinates in a highly nonlinear way. 
Therefore, to separate the vibron degrees of freedom from the phonon coordinates, a mean field procedure is applied. The full Hamiltonian $\hat{H}$ is thus written as
$\hat{H}=\hat{H}_{eff}+H_{p}+\Delta H $, where $H_{p}$ is the phonon Hamiltonian and where $\hat{H}_{eff}=<\hat{H}>-H_{p}$ denotes the effective Hamiltonian of the dressed anharmonic vibrons. $\Delta H =\hat{H}-<\hat{H}>$ stands for the remaining part of the vibron-phonon interaction. The symbol $<...>$ represents a thermal average over the phonon degrees of freedom at temperature $T$.

\subsection{The effective vibron Hamiltonian and the two-vibron eigenstates}

The effective dressed anharmonic vibron Hamiltonian is written as
\begin{eqnarray}
&&\hat{H}_{eff}=\sum_{n} \hat{\omega}_{0}b^{+}_{n}b_{n}-\hat{A}b^{+2}_{n}b_{n}^{2}-\hat{B}b^{+}_{n+1}b^{+}_{n}b_{n+1}b_{n}  \nonumber \\
&&-J_{1}[\Phi(N_{n}+N_{n+1})b^{+}_{n}b_{n+1}+H.c.] \nonumber \\
&&-J_{2}[\Phi(N_{n}+N_{n+1})^{4}b^{+2}_{n}b_{n+1}^{2}+H.c.] \nonumber \\
&&-J_{3}[\Phi(N_{n}+N_{n+1})b^{+}_{n}[N_{n}+N_{n+1}]b_{n+1}+H.c.] 
\label{eq:HHATEFF}
\end{eqnarray}
where $\Phi(X)=exp(-S(T)[1+2\eta+2\eta X])$ and where $S(T)$ is the coupling constant 
introduced by Ivic and co-workers as ($k_{B}$ denotes the Boltzmann constant)
\begin{equation}
S(T)=\frac{4E_{B}}{N\Omega_{c}} \sum_{q} \mid sin(\frac{q}{2}) \mid cos(\frac{q}{2})^{2} coth(\frac{\Omega_{q}}{2k_{B}T})
\label{eq:S}
\end{equation}

In Ref. \cite{kn:pouthier4}, a detailed analysis of the two-vibron energy spectrum of the  
Hamiltonian $\hat{H}_{eff}$ is presented. Within the number state method \cite{kn:scott,kn:pouthier1,kn:pouthier2,kn:pouthier3}, the two-vibron wave function is first expanded as $\mid \Psi \rangle = \sum \Psi(n_{1},n_{2}) \mid n_{1},n_{2} ) $ where $\mid n_{1},n_{2} )$ denotes a local basis vector characterizing two vibrons located onto the sites $n_{1}$ and $n_{2}$, respectively. Note that the restriction $n_{2} \geq n_{1}$ is applied due to the indistinguishable character of the vibrons so that the dimension of the two-vibron subspace is $N(N+1)/2$. Then, by taking advantage of the lattice periodicity, the wave function is expanded as a Bloch wave as
\begin{equation}
\Psi(n_{1},n_{2}=n_{1}+m)= \frac{1}{\sqrt{N}} \sum_{n_{1}} e^{ik(n_{1}+m/2)}\Psi_{k}(m)
\label{eq:bloch}
\end{equation}
where the total momentum $k$ is associated to the motion of the center of mass of the two vibrons whereas the resulting wave function $\Psi_{k}(m)$ refers to the degree of freedom $m$ which characterizes the distance between the two vibrons. 
Since $k$ is a good quantum number, the Hamiltonian $\hat{H}_{eff}$ appears as block diagonal and the Schrodinger equation can be solved for each $k$ value. For a given $k$ value, the protein exhibits $(N+1)/2$ eigenstates $\mid \Psi_{k\sigma} \rangle$, where the index $\sigma=1,...,(N+1)/2$ refers to the position of the states along the energy axis.  Due to the nonlinear sources, there are two different eigenstates, i.e. the two-vibron free states (TVFS) and the TVBS. The TVFS correspond to a delocalization of the separating distance $m$ between the two vibrons. The wave function $\Psi_{k}(m)$ behaves as a plane wave and the TVFS belong to an energy continuum. By contrast, TVBS correspond to a localization of the separating distance between the two vibrons and characterize the trapping of the two quanta over only a few neighboring sites. We have shown that the protein supports two kinds of bound states, called TVBS-I and TVBS-II, respectively. The TVBS-I, denoted $\mid \Psi_{k,\sigma=1} \rangle$, are located below the TVFS continuum over the entire Brillouin zone whereas for TVBS-II, two situations occur depending on the strength of the small polaron binding energy. For small values of $E_{B}$, the band disappears inside the continuum when $\mid k \mid$ is lower than a critical wave vector $k_{c}$ whereas, for strong values of $E_{B}$, the band is located below the continuum over the entire Brillouin zone. As a result, the notation $\mid \Psi_{k,\sigma=2} \rangle$ refers either to a free state or to TVBS-II, depending on the situation. In the harmonic situation, both TVBS-I and TVBS-II appear as combinations of states involving the trapping of the two vibrons onto the same amide-I mode and onto nearest neighbor amide-I modes. By contrast, the intramolecular anharmonicity reduces the hybridization between these two kinds of trapping so that TVBS-I refers to the trapping of the two vibrons onto the same amide-I mode whereas TVBS-II characterizes the trapping onto nearest neighbor amide-I vibrations.

\subsection{The vibron-phonon coupling Hamiltonian}

By comparing Eq.(\ref{eq:HHAT}) and Eq.(\ref{eq:HHATEFF}), it is straightforward to show that the coupling Hamiltonian $\Delta H$ corresponds to a modulation of the different lateral contributions describing vibron hops, i.e. the terms proportional to $J_{1}$, $J_{2}$ and $J_{3}$ in Eq.(\ref{eq:HHAT}). However, in $\alpha$-helix proteins, it has been shown that $J_{2} \approx J_{3} \approx J_{1}/\omega_{0}$ \cite{kn:pouthier4}. As a result $J_{2}$ and  $J_{3}$ are of about three orders of magnitude lesser than $J_{1}$ and can be neglected. The coupling Hamiltonian $\Delta H$ is thus written as
\begin{eqnarray}
\Delta H &=& -J_{1} \sum_{n,\delta=\pm 1} [\Theta^{+}_{n}(N_{n}+1)\Theta_{n+\delta}(N_{n+\delta})- \nonumber \\
& & \left\langle \Theta^{+}_{n}(N_{n}+1)\Theta_{n+\delta}(N_{n+\delta}) \right\rangle ]b^{+}_{n}b_{n+\delta}
\end{eqnarray}
In addition, the small polaron binding energy is of about one order of magnitude smaller than the phonon cutoff frequency so that the dressing operator Eq.(\ref{eq:theta}) can be linearized \cite{kn:ivic3,kn:ivic4}. As a consequence, by neglecting the rather small parameter $\eta$ in Eq.(\ref{eq:theta}) \cite{kn:pouthier4}, the coupling between the anharmonic polarons and the remaining phonons is finally expressed as     
\begin{equation}
\Delta H \approx -\sum_{n,\delta=\pm 1} \Delta J(n,n+\delta)b^{+}_{n}b_{n+\delta}
\label{eq:DH}
\end{equation}
where 
\begin{eqnarray}
&&\Delta J(n,n+\delta)=  \\
&&J_{1}\sum_{q} -i\sqrt{\frac{E_{B}}{2N\Omega_{q}}} \frac{sin(q)e^{-iqn}}{\mid sin(q/2)\mid}(1-e^{-iq\delta})a^{+}_{q}-H.c. \nonumber
\label{eq:DH1}
\end{eqnarray}
Therefore, within the anharmonic polaron point of view, the main contribution of the coupling with the remaining phonons corresponds to a random modulation of the single-vibron hopping constant. As shown in the following sections, this coupling is responsible for the relaxation of the two-vibron eigenstates.  

\section{Two-vibron bound state relaxation rate}

Due to the coupling Hamiltonian $\Delta H$ (Eq.(\ref{eq:DH})), TVBS do not represent exact eigenstates of the whole polaron-phonon system. More precisely, this coupling is responsible for the occurrence of transitions between two-vibron states mediated by the emission or the absorption of acoustic phonons.  
Therefore, by using the Golden rule formula, the rate for the transition from a TVBS $\mid \Psi_{k\sigma_b} \rangle$ with frequency $\omega_{k\sigma_b}$ to another state $\mid \Psi_{k'\sigma'} \rangle$\ with frequency $\omega_{k'\sigma'}$ is expressed as 
\begin{eqnarray}
W_{k\sigma_b \rightarrow k'\sigma'}&=& 2 \pi
\sum_{\alpha,\beta} P_{\alpha} \mid \langle \Psi_{k\sigma_b},\alpha \mid \Delta H \mid
\Psi_{k'\sigma'},\beta \rangle \mid^{2} \nonumber \\ 
& &\delta(\omega_{k\sigma_b}+\Omega_{\alpha}-\omega_{k'\sigma'}-\Omega_{\beta})  
\label{eq:rate1}
\end{eqnarray}
Eq.(\ref{eq:rate1}) describes a transition in the course of which the phonon bath evolves from an initial state $\mid \alpha \rangle$ with frequency $\Omega_{\alpha}$ to a final state $\mid \beta \rangle$ with frequency $\Omega_{\beta}$. Since the bath is assumed to be in thermal equilibrium at temperature $T$, a statistics is realized over the initial state with the probability occupation $P_{\alpha}$ and a sum over all the possible final bath states is performed. By inserting the expression of the coupling Hamiltonian $\Delta H$ (Eq.(\ref{eq:DH})), the total rate for leaving the state $\mid \Psi_{k\sigma_b} \rangle$ obtained by summing over all possible transitions is expressed as
\begin{eqnarray}
& &W_{k\sigma_b} = 2Re \sum_{n,\delta} \sum_{n',\delta'} \sum_{k',\sigma'}\int_{0}^{\infty} dt e^{i(\omega_{k\sigma_b}-\omega_{k'\sigma'})t} \nonumber \\
& & \langle \Psi_{k\sigma_b} \mid b_{n}^{+} b_{n+\delta} \mid \Psi_{k'\sigma'} \rangle \langle \Psi_{k'\sigma'} \mid b_{n'}^{+} b_{n'+\delta'} \mid  \Psi_{k\sigma_b} \rangle\nonumber \\ 
& & \langle \Delta J(n,n+\delta,t) \Delta J(n',n'+\delta',0) \rangle 
\label{eq:rate2}
\end{eqnarray}
where $\langle ... \rangle$ stands for an average over the phonon bath and 
where the operators $\Delta J$ depend on time $t$ according to an Heisenberg 
representation with respect to the phonon Hamiltonian $H_{p}$. 

As shown in Eq.(\ref{eq:rate2}), the TVBS relaxation rate is expressed in terms of the Fourier transform of the correlation function of the coupling $\Delta J $. 
The characteristic time of this rate is the correlation time $\tau_{c} $ of the phonon bath which corresponds to the time for which the correlation functions vanish. In a general way, $\tau_{c}$ is about 1 ps for phonons in low-dimensional molecular lattices \cite{kn:pouthier2}. We thus assume that this correlation time is sufficiently small in order to neglect the spatial correlations in the phonon bath. As a result, the correlation functions of the coupling $\Delta J$ which appear in Eq.(\ref{eq:rate2}) are nonzero if $n=n'+\delta'$ and $n'=n+\delta$, only. Therefore, by performing both the time integration as well as the thermal average in  Eq.(\ref{eq:rate2}), the relaxation rate is finally expressed as
\begin{eqnarray}
W_{k\sigma_b}&=& \frac{32J_{1}^{2}E_{B}}{\Omega_{c}^{2}} \sum_{k'\sigma'} 
Z_{k\sigma_b \rightarrow k'\sigma'} \nonumber \\
&& [F(\omega_{k'\sigma'}-\omega_{k\sigma_b}) n(\omega_{k'\sigma'}-\omega_{k\sigma_b})+ \nonumber \\ 
&&F(\omega_{k\sigma_b}-\omega_{k'\sigma'}) (1+n(\omega_{k\sigma_b}-\omega_{k'\sigma'}))]
\label{eq:rate3}
\end{eqnarray}
where $n(\Omega)$ denotes the Bose-Einstein phonon distribution at temperature $T$ and where the coupling distribution $F(\Omega)$, which measures the probability for the exchange of a phonon with frequency $\Omega$ during the process, is defined as  
\begin{eqnarray}
F(\Omega)= \left\{ \begin{array}{cc} 
\frac{\Omega}{\Omega_{c}} \sqrt{1-(\frac{\Omega}{\Omega_{c}})^{2}} & \mbox{if $\Omega>0$} \\
0 & \mbox{if $\Omega<0$}
\end{array}
\right.
\label{eq:F}
\end{eqnarray}
In Eq.(\ref{eq:rate3}), $Z_{k\sigma \rightarrow k'\sigma'}$ characterizes the strength of the coupling between the two-vibron eigenstates $\mid \Psi_{k\sigma} \rangle$ and $\mid \Psi_{k'\sigma'} \rangle$\ due to the vibron-phonon interaction. This coupling is expressed as
\begin{equation}
Z_{k\sigma \rightarrow k'\sigma'}=\sum_{n,\delta=\pm1} \mid\langle \Psi_{k\sigma} \mid b_{n}^{+} b_{n+\delta} \mid \Psi_{k'\sigma'} \rangle \mid^{2}
\label{eq:Z}
\end{equation}
After some algebraic manipulations, $Z_{k\sigma \rightarrow k'\sigma'}$ is expressed in terms of the wave functions as
\begin{eqnarray}
&&Z_{k\sigma \rightarrow k'\sigma'}=
 \frac{1}{N}\mid \sum_{m} \frac{\Psi_{k\sigma}(m)}{\Delta(m)}( \frac{\Psi_{k'\sigma'}(m-1)}{\Delta(m-1)} e^{i(k'-k)m/2} \nonumber \\
&&+\frac{\Psi_{k'\sigma'}(m+1)}{\Delta(m+1)} e^{-i(k'-k)m/2} \mid^{2} \nonumber \\
& &+\frac{1}{N}\mid \sum_{m} \frac{\Psi_{k'\sigma'}(m)}{\Delta(m)}( \frac{\Psi_{k\sigma}(m-1)}{\Delta(m-1)} e^{-i(k'-k)m/2} \nonumber \\
&&+\frac{\Psi_{k\sigma}(m+1)}{\Delta(m+1)} e^{i(k'-k)m/2} \mid^{2}
\label{eq:Z1}
\end{eqnarray}
where the convention $\Psi_{k\sigma}(-1)$ is used.

As shown in Eq.(\ref{eq:rate3}), the rate depends on the temperature through the average number of phonons. Moreover, the temperature is involved in the definition of the two-vibron wave functions due to the dressing effect \cite{kn:pouthier4}. The relaxation rate exhibits two contributions connected to the absorption (term proportional to $n(\Omega)$) and to the emission (term proportional to $1+n(\Omega)$) of an acoustic phonon, respectively. Note that Eq.(\ref{eq:rate3}) clearly shows that the relaxation rate $W_{k\sigma_b}$ is expressed as the sum over the rate connected to the different relaxation channels $W_{k\sigma_b \rightarrow k'\sigma'}$. As a result, each channel can be characterized separately. 

At this step, the diagonalization of the Hamiltonian $H_{eff}$ realized in Ref.\cite{kn:pouthier4} allows us to calculate both the two-vibron eigenstates $\mid \Psi_{k\sigma} \rangle $ and eigenenergies $\omega_{k\sigma}$. Then, by using Eqs.(\ref{eq:rate3})-(\ref{eq:Z}), the TVBS relaxation rate can be computed. This procedure is illustrated in the following section. 

\section{Numerical results}

In this section, the previous formalism is applied to compute the TVBS relaxation rate in an anharmonic $\alpha$-helix protein. The intramolecular anharmonicity is described by a single parameter, namely the anharmonic constant $A$, which ranges between $0$ and $10$ cm$^{-1}$ \cite{kn:pouthier4,kn:hamm1,kn:hamm2}. The small polaron binding energy $E_{B}$ is taken as a parameter which extends from 0 to 15 cm$^{-1}$. 
The phonon cutoff frequency $\Omega_{c}$ is fixed to 100 cm$^{-1}$ and the hopping constant is set to $J=8$ cm$^{-1}$. 

The temperature dependence of the zero wave vector TVBS-I relaxation rate (full circles) is shown in Figs. 1a and 1b for two typical situations. The empty circles correspond to the rate for the relaxation over all the other TVBS-I whereas the empty squares represent the rate for the decay into the set of the second eigenstates $\mid \Psi_{k\sigma=2} \rangle$. Note that as remained in Sec. II.B., such states refer to either free or bound states (TVBS-II), depending on the nonlinearity. 

When A=8 cm$^{-1}$ and $E_{B}=4$ cm$^{-1}$ (Fig. 1a), the relaxation rate exhibits a quasi-linear dependence versus the temperature, excepted at low temperature. More precisely, the linear regime is reached when the temperature is greater than 50 K whereas the rate shows a power law dependence at low temperature. The relaxation rate is equal to $0.049$ cm$^{-1}$ at $T=5$ K and reaches 7.64 cm$^{-1}$ at $T=315$ K. At low temperature, Fig. 1a clearly indicates that the main mechanism for the relaxation involves the decay of the zero wave vector TVBS-I into the other TVBS-I. For instance, this channel represents $99.8$ $\%$ of the relaxation at $T=5$ K. As increasing the temperature, the relaxation over the other TVBS-I decreases and the rate for the relaxation over the states $\mid \Psi_{k\sigma=2} \rangle$ increases very slightly. Indeed, at $T=315$ K, the relaxation over the other TVBS-I represents 55.54 $\%$. However, the second channel, i.e. the relaxation over all the states $\sigma=2$, represents only 7.22 $\%$ which indicates that 37.24 $\%$ of the relaxation involves the decay of the TVBS-I into the TVFS continuum. 

When $E_{B}=12$ cm$^{-1}$ (Fig. 1b), the relaxation rate for the zero wave vector TVBS-I exhibits almost the same temperature dependence as in the previous case. Nevertheless, the rate is more important since it is equal to $0.19$ cm$^{-1}$ at $T=5$ K and reaches $19.58$ cm$^{-1}$ at $T=315$ K. However, the main difference with the previous case originates in the nature of the relaxation channels. Indeed, although the relaxation over the other TVBS-I remains the main pathway at low temperature, this is no longer true at high temperature. Indeed, as shown in Fig. 1b,
the decay into the other TVBS-I represents almost 100 $\%$ of the relaxation at $T=5$ K. However, as increasing the temperature, the relaxation over the other TVBS-I strongly decreases whereas the rate for the relaxation over the states $\mid \Psi_{k\sigma=2} \rangle$ increases and becomes the dominate contribution (the transition occurs around 130 K). At high temperature, i.e. $T=315$ K, the relaxation according to the second channel represents 85.60 $\%$ whereas the decay into the other TVBS-I charaterizes 10.84 $\%$
of the global rate. As a consequence, for such a strong nonlinear situation, the decay of the TVBS-I into the TVFS is no more than 5 $\%$ at high temperature. 

The behavior of the TVBS-I relaxation rate as a function of the anharmonicity is displayed in Figs. 2. The calculations are performed at $T=310$ K and for three different values of the small polaron binding energy. When $E_{B}=4$ cm$^{-1}$ (Fig. 2a), the rate slightly decreases as the anharmonicity increases. It is equal to $8.28$ cm$^{-1}$ when $A=0$ and to $7.26$ cm$^{-1}$ when $A=10$ 
cm$^{-1}$. In marked contrast, the rate for the decay into the other TVBS-I first increases to reach a maximum equal 6.02 cm$^{-1}$ when $A=2$ cm$^{-1}$. Then, it decreases and is equal to 3.48 cm$^{-1}$ when A=10 cm$^{-1}$. As shown in Fig. 2a, the rate for the decay into the second eigenstates $\mid \Psi_{k\sigma=2} \rangle$ is rather small whatever the anharmonicity although it increases when A increases. Consequently, in this low nonlinear regime ($E_{B}=4$ cm$^{-1}$), the main part of the relaxation of TVBS-I involves the decay into both the other TVBS-I and the TVFS continuum.

As when increasing the small polaron binding energy, (Fig. 2b and 2c), the TVBS-I relaxation rate behaves in a similar way with respect to the anharmonicity and slightly decreases as A increases. However, the rate increases with $E_{B}$ since it is equal to 16.31 cm$^{-1}$ when $E_{B}=8$ cm$^{-1}$ and $A=0$ (Fig. 2b) and reaches 22.73 cm$^{-1}$ when $E_{B}=12$ cm$^{-1}$ and $A=0$ (Fig. 2c). In a marked contrast, the relaxation pathways are strongly modified when the small polaron binding energy is increased. Indeed, when $E_{B}=8$ cm$^{-1}$ (Fig. 2b), the rate for the decay into the other TVBS-I decreases as A increases. By contrast, the rate for the decay into the second eigenstates $\mid \Psi_{k\sigma=2} \rangle$ increases. This second channel becomes slightly more efficient than the first channel when the anharmonicity exceeds 7 cm$^{-1}$. When $A=10$ cm$^{-1}$, the first channel represents 21.92 $\%$ of the relaxation whereas the second channel characterizes 43.31 $\%$ of the decay. Such a behavior appears more pronounced when the small binding energy is set to $E_{B}=12$ cm$^{-1}$. In that case, Fig. 2c clearly shows that the decay of the TVBS-I into the TVFS continuum is rather weak. Its contribution is less than 5 $\%$ when A is greater than 3.5 cm$^{-1}$. For a small anharmonicity, the rates for the decay into the other TVBS-I and into the second eigenstates are of the same order of magnitude. However, as increasing the anharmonicity, the second channel becomes dominant since the corresponding rate represents almost 90 $\%$ of the global rate when A=10 cm$^{-1}$.

The dependence of the TVBS-I relaxation rate on the small polaron binding energy is shown  
in Fig. 3 for $T=310$ K, and $A=8$ cm$^{-1}$. The global rate evolves in a quasi-linear way and varies from 1.98 cm$^{-1}$ when $E_{B}=1$ cm$^{-1}$ to 23.19 cm$^{-1}$ when $E_{B}=15 $ cm$^{-1}$. As shown in Fig. 3, the more surprising results correspond to the behavior of the rates connected to the first and to the second channel. Indeed, for small $E_{B}$ values, the decay into the other TVBS-I is the dominant relaxation pathway. For instance, when $E_{B}=2$ cm$^{-1}$, this first channel represents 67.66 $\%$ of the relaxation whereas the contribution of the second channel is 2.13 $\%$. Therefore, 30.21 $\%$ of the relaxation involve the decay of the TVBS-I into the TVFS continuum. However, as $E_{B}$ increases, the rate connected to the second channel increases and becomes the main contribution for strong $E_{B}$ values. When $E_{B}=14$ cm$^{-1}$, the second channel represents 91.54 $\%$ of the relaxation whereas the contribution of the first channel is 6.81 $\%$. Note that both channels contribute in a similar way
when $E_{B}$ is about $7.5$ cm$^{-1}$. 

The correlation between the relaxation channels and the nature of the two-vibron eigenstates, is illustrated in Figs. 4 for $T=310$ K and $A=8$ cm$^{-1}$. The upper panel represents the corresponding two-vibron energy spectrum whereas the lower panel displays the wave vector dependence of the relaxation rates. More precisely, open circles characterize the rate $W_{0,1 \rightarrow k\sigma=1}$ for the decay of the zero wave vector TVBS-I into the TVBS-I with wave vector $k$. In the same way, open squares correspond to the decay of the zero wave vector TVBS-I into the state $\mid \Psi_{k\sigma=2} \rangle $, i.e. $W_{0,1 \rightarrow k\sigma=2}$.

When $E_{B}=4$ cm$^{-1}$ (Fig. 4a), the TVBS-II band occurs at the end of the first Brillouin zone, only. For the first relaxation channel, i.e. the relaxation over the other TVBS-I, the rate decreases as the modulus of the wave vector increases. In other words, the decay into low wave vector TVBS-I is the dominant relaxation pathway. Although such an effect is not correlated to the energy spectrum, this is no longer true for the $k$ dependence of the rate connected to the second channel. Indeed, Fig. 4a clearly shows that the second channel opens at the end of the first Brillouin zone, only, i.e. when the TVBS-II band occurs. Therefore, this second channel becomes the dominant relaxation pathway at the end of the Brillouin zone. When $E_{B}=10$ cm$^{-1}$ (Fig. 4b), the results are slightly different since $E_{B}$ is strong enough so that the TVBS-II band is localized below the continuum over the entire Brillouin zone. As a result, the second channel is clearly the main mechanism for the relaxation whatever the value of the wave vector.   

In Fig. 5, the influence of the small polaron binding energy on the relaxation rate of the $\mid \Psi_{k=0,\sigma=2} \rangle$ eigenstate is shown at biological temperature (T=310 K) and for $A=8$ cm$^{-1}$. As in the previous figures, full circles characterize the global rate, empty circles represent the rate for the decay into the TVBS-I and empty squares correspond to the rate for the decay into all the other $\mid \Psi_{k,\sigma=2} \rangle$ eigenstates. In addition, empty triangles characterize the rate for the decay into the TVFS continuum. The figure clearly shows that the system exhibits two regimes depending on the value of $E_{B}$. 

For the small values of $E_{B}$, i.e. typically $E_{B}< 8.5$ cm$^{-1}$, the rate increases in a quasi-linear way with $E_{B}$. It is equal to 1.65 cm$^{-1}$ when $E_{B}=1 $ cm$^{-1}$ and reaches 13.67 cm$^{-1}$ when $E_{B}=8.0 $ cm$^{-1}$. As shown in Fig. 5, the relaxation into both the TVBS-I and the other $\mid \Psi_{k,\sigma=2} \rangle$ eigenstates can be neglected. In other words, the main relaxation pathway corresponds to the decay of the $\mid \Psi_{0,\sigma=2} \rangle$ eigenstate into the TVFS continuum. When $E_{B}$ becomes greater than a critical value, the global rate behaves in a different manner with respect to the small polaron binding energy. It increases more rapidly than the previous linear regime to reach 38.75 cm$^{-1}$ for $E_{B}=15$ cm$^{-1}$. Moreover, the rate for the relaxation into the TVBS-I increases with $E_{B}$ whereas the rate for the relaxation into TVFS slightly decreases. For strong $E_{B}$, the decay into TVBS-I represents 64 $\%$ of the relaxation whereas the decay into TVFS corresponds to 33 $\%$. Note that although the rate for the decay into the  $\mid \Psi_{k,\sigma=2} \rangle$ eigenstates sligtly increases around the transition, it decreases as $E_{B}$ increases and can be neglected for a strong nonlinearity.

Finally, the relaxation rate of the $\mid \Psi_{k=0,\sigma=2} \rangle$ eigenstate versus the anharmonicity is shown in Fig. 6 for T=310 K and for $E_{B}=10$ cm$^{-1}$. As in Fig. 5, the different rates exhibit two regimes depending on the anharmonicity.
For a small anharmonicity, i.e. $A<4$ cm$^{-1}$, the global rate is almost independent on the anharmonic parameter and it is equal to 17.00 cm$^{-1}$. In addition, the decay into the TVFS continuum appears as the main mechanism for the relaxation. In a marked contrast, for a strong anharmonicity, i.e. $A>4$ cm$^{-1}$, the rate increases as the anharmonic parameter increases and reaches 23.19 cm$^{-1}$ when $A=10$ cm$^{-1}$. The rate for the relaxation over the TVFS continuum decreases whereas the rate for the decay into the TVBS-I increases. This latter rate becomes the most important when $A=10$ cm$^{-1}$ so that the decay into TVBS-I represents almost 50 $\%$ of the relaxation.

\section{Discussion}

To discuss and interpret the previous numerical results, let us first consider the behavior of the TVBS relaxation rate at low temperature. Since the zero wave vector TVBS-I lies at the bottom of the two-vibron energy spectrum, its decay involves phonon absorption, only. The TVBS-I relaxation rate is thus proportional to the Bose-Einstein distribution which selects the frequency range of the phonons which are exchanged. In that context, transitions involving low frequency phonons takes place at low temperature. Therefore, when the thermal energy $k_{B}T$ is lower than the energy gap between bound and free states, the TVBS-I can just decays into the other TVBS-I, as shown in Fig. 1. Note that phonon emission participates in the decay of the other two-vibron states so that the corresponding rate reaches a finite value at zero temperature (not considered in the numerical analysis). As a consequence, the low temperature behavior of the rate does not depend on the system nonlinearity and essentially originates in the shape of the phonon distribution.

This is no longer the case at biological temperature for which an approximate expression of the relaxation rate can be determined. To proceed, we assume that the two-vibron bandwidth is smaller when compared with both the thermal energy $k_{B}T$ and the phonon cutoff energy $\hbar\Omega_c$. Therefore, the Bose-Einstein distribution can be linearized according to the temperature and the distribution function F($\Omega$) (Eq.(\ref{eq:F})) can be written as 
$F(\Omega)\approx\Omega/\Omega_{c}$. As a result, the relaxation rate Eq.(\ref{eq:rate3}) connected to the two-vibron eigenstate $\mid \Psi_{k\sigma} \rangle $ can be approximated as
\begin{equation}
W_{k\sigma}\approx \frac{32J^{2}_{1}E_{B}k_{B}T}{\Omega_{c}^{3}}\sum_{k'\sigma'}Z_{k\sigma \rightarrow k'\sigma'}
\label{eq:Wapprox}
\end{equation}
At this step, Eq.(\ref{eq:Wapprox}) can be simplified because the sum over $k'\sigma'$  leads to the occurrence of the closure relation. Therefore, by using the following identity  
\begin{equation}
\langle \Psi_{k\sigma} \mid b^{+}_{n}b_{n} b^{+}_{n'}b_{n'}\mid \Psi_{k\sigma} \rangle = \frac{1}{N}\mid\Psi_{k\sigma}(\mid n-n'\mid)\mid^{2}
\label{eq:corr}
\end{equation}
the relaxation rate is finally expressed as 
\begin{equation}
W_{k\sigma} \approx\frac{128J^{2}_{1}E_{B}k_{B}T}{\Omega_{c}^{3}}(1+\frac{1}{2}\mid\Psi_{k\sigma}(1)\mid^{2})
\label{eq:Wapprox1}
\end{equation}

Eq.(\ref{eq:Wapprox1}) yields a rather good approximation which allows us to interpret and to understand the numerical results. 
First of all, it accounts for the observed temperature dependence of the relaxation rate. Indeed, Eq.(\ref{eq:Wapprox1}) clearly shows that the rate increases in a linear way as the temperature increases, in a perfect agreement with the numerical results displayed in Figs. 1. This feature originates in the linearization of the Bose-Einstein distribution. Note that the wave function $\Psi_{k,\sigma}(m)$ depends on the temperature in a complicated manner through the dressing effect \cite{kn:pouthier4}. However, our results indicate that such a dependence remains rather small when compared with the influence of the Bose-Einstein factor.

Then, the wave function dependence of the relaxation rate allows us to understand the influence of the intramolecular anharmonicity as shown in Figs. 2 and 6.  
For TVBS-I, Figs. 2 clearly show that the rate decreases as the anharmonicity increases. In fact, since TVBS-I refer to the trapping of the two vibrons onto the same amide-I vibration, the wave function $\Psi_{k\sigma=1}(m)$ is maximum for $m=0$ and decreases with $m$ according to a quasi-exponential way \cite{kn:pouthier4}. As a consequence, when the anharmonicity increases, the trapping process is enhanced so that the extension of the wave function around $m=0$ is reduced. Therefore, $\mid \Psi_{k\sigma=1}(0) \mid^{2}$ increases whereas $\mid \Psi_{k\sigma=1}(1) \mid^{2}$, as the relaxation rate, decreases.
In Fig. 6, the anharmonicity dependence of the rate for the $\mid \Psi_{k=0\sigma=2}\rangle$ eigenstate exhibits two regimes which originates in the nature of the state itself. Indeed, as pointed out in Sec. II. B., $\mid \Psi_{k=0\sigma=2}\rangle$ refers either to a free or to a bound state (TVBS-II), depending on the nonlinearity. Such a behavior is displayed in Fig. 6 since the change of regime corresponds to the transition from a TVFS to a TVBS-II. As shown in Ref. \cite{kn:pouthier4}, $\mid \Psi_{k=0\sigma=2}\rangle$ corresponds to a TVFS for a small anharmonicity. In that case, the probability $\mid \Psi_{k\sigma=2}(m) \mid^{2}$ is almost independent on the anharmonicity and scales as $1/N$. Consequently, the relaxation rate does not depend on $A$, as shown in Fig. 6. By contrast, as when increasing the anharmonicity, the state $\mid \Psi_{k=0\sigma=2}\rangle$ becomes bounded. It refers to a TVBS-II which characterizes the trapping of the two vibrons onto two nearest neighbor amide-I vibrations. Therefore, when the anharmonicity increases, the trapping process is enhanced so that the wave function $\Psi_{k\sigma=2}(m)$ tends to localize around $m=1$. The corresponding rate suddenly increases with $A$ as clearly shown in Fig. 6.

Finally, the approximated expression of the rate gives a comprehensive explanation of the influence of the small polaron binding energy (see Figs. 3 and 5). Indeed, Eq.(\ref{eq:Wapprox1}) clearly shows that the dependence of the rate with respect to $E_{B}$ is twofold. First, the rate depends linearly on $E_{B}$. This feature originates in the fact that the rate is proportional to the intensity of the coupling between the dressed anharmonic vibrons and the remaining phonons (see Eq.(\ref{eq:rate1})). Then, the $E_{B}$ dependence is included in the wave function dependence of the rate. 
As when increasing the small polaron binding energy, the trapping process involved in the formation of the TVBS-I is enhanced so that $\Psi_{k\sigma=1}(1)$ decreases. Therefore, the linear evolution of the TVBS-I relaxation rate with respect to $E_{B}$ is slightly damped, i.e. the rate evolves more slowly than the corresponding linear law (see Fig. 3).
As previously, Fig. 5 clearly shows that the nature of the state $\mid \Psi_{k=0\sigma=2}\rangle$ exhibits a transition as a function of the small polaron binding energy. For small $E_{B}$ values, this state is a TVFS which the wave function does not significantly depend on the nonlinearity. Therefore, the rate evolves in a linearly way versus $E_{B}$. By contrast, when the small polaron binding energy is sufficiently important, this state becomes a TVBS-II \cite{kn:pouthier4}. As a consequence, the trapping onto two nearest neighbor sites is enhanced so that $\Psi_{k\sigma=2}(1)$ increases to reach unity. The corresponding relaxation rate suddenly evolves more rapidly than the previous linear law (see Fig. 5).  

As shown in Eq.(\ref{eq:Wapprox1}), the dependence of the relaxation rate on the nature of the corresponding two-vibron state arises from the factor $(1+\mid\Psi_{k\sigma}(1)\mid^{2}/2)$. As a consequence, the global rate does not depend significantly on the fact that the state is a TVFS or a TVBS. In other words, the lifetime of both kinds of states is about the same order of magnitude.
Indeed, $\mid\Psi_{k\sigma}(1)\mid^{2}$ refers to the probability to find the two vibrons onto two nearest neighbor sites. In a TVFS, this probability scales as $1/N$ and vanishes asymptotically as $N$ tends to infinity. For a TVBS, such a probability depends on the nature of the bound between the two vibrons and ranges between $0$ and $1$. For instance, for a strong nonlinearity, i.e. for strong values of both $A$ and $E_{B}$, TVBS-I refer to the trapping of the two vibrons onto the same amide-I vibration whereas TVBS-II characterize the trapping onto two nearest neighbor sites. Therefore, $\mid\Psi_{k\sigma}(1)\mid^{2}$ vanishes for TVBS-I whereas it is equal to unity for TVBS-II. In that context, if we define $W=128J^{2}_{1}E_{B}k_{B}T/\Omega_{c}^{3}$, the relaxation rates are expressed as
$W_{TVBS-I} \approx W_{TVFS} \approx W $ and $W_{TVBS-II}\approx  3W/2 $.

In a marked contrast, the nature of the relaxation channels drastically depends on the characteristic of the two-vibron states involved in the process. Such effects originate in the coupling between the two-vibron eigenstates mediated by the vibron-phonon interaction and characterized by the constant $Z_{k\sigma \rightarrow k'\sigma'}$ (see Eqs.(\ref{eq:Z})-(\ref{eq:Z1})). As shown in the definition of the coupling Hamiltonian $\Delta H$ (Eq.(\ref{eq:DH})), this interaction characterizes transitions between two-vibron states in the course of which a single vibron is transfered from a given site to its nearest neighbor. 

To understand this feature, let us first consider the relaxation channels connected to the decay of the TVBS-I. Within the strong nonlinear limit, the TVBS-I wave function is essentially localized around $m=0$ so that $\Psi_{k\sigma=1}(m)$ is almost equal to $\delta_{m,0}$. Therefore, it is straightforward to show that the coupling constant can be approximated as $Z_{k\sigma=1 \rightarrow k'\sigma'} \approx 4\mid \Psi_{k\sigma=1}(0)\mid^{2}\mid \Psi_{k'\sigma'}(1)\mid^{2}/N$. The corresponding relaxation rate is thus written as
\begin{equation}
W_{k\sigma=1\rightarrow k'\sigma'} \approx\frac{W}{N}\mid \Psi_{k\sigma=1}(0)\mid^{2}\mid \Psi_{k'\sigma'}(1)\mid^{2}
\label{eq:WTVBSI}
\end{equation}
The relaxation channel for the TVBS-I corresponds to the decay into states which favor the trapping of the two vibrons onto two nearest neighbor amide-I vibrations, i.e. states for which $\mid \Psi_{k'\sigma'}(1)\mid^{2}$ is maximum.  
Within the strong nonlinear limit, such states refer to the TVBS-II. As a consequence, in a perfect agreement with the numerical results displayed in Figs. 1a, 2c, 3 and 4b, Eq.(\ref{eq:WTVBSI}) clearly shows that the decay into TVBS-II is the main mechanism for the TVBS-I relaxation. Note that both $\mid \Psi_{k\sigma=1}(0)\mid^{2}$ and $\mid \Psi_{k\sigma=2}(1)\mid^{2}$ are almost $k$ independent so that the rate for the decay over all the TVBS-II states is equal to $W$. However, when the nonlinearity slightly decreases, i.e. when either (or both) $A$ and $E_{B}$ decreases, the TVBS-I wave function delocalizes so that $\Psi_{k\sigma=1}(0)$ decreases and, in the same time, $\Psi_{k\sigma=1}(1)$ increases. As a result, the coupling between a given TVBS-I and the other TVBS-I is turned on which opens the corresponding relaxation channel. In the same way, the TVBS-II wave function extends itself around $m=1$ leading to the decrease of $\Psi_{k\sigma=2}(1)$. As a consequence, the coupling between TVBS-I and TVBS-II decreases, as observed in the numerical results. 

Within the strong nonlinear limit, the TVBS-II wave function is essentially localized around $m=1$ so that $\Psi_{k\sigma=2}(m)$ is almost equal to $\delta_{m,1}$. Therefore, by following the same procedure as in the previous paragraph, it is straightforward to show that the coupling constant is 
$Z_{k\sigma=2 \rightarrow k'\sigma'} \approx 4\mid \Psi_{k\sigma=2}(1)\mid^{2}
\mid \Psi_{k'\sigma'}(0)+\Psi_{k'\sigma'}(2)\exp(i(k-k'))/\sqrt{2}\mid^{2}/N$. This expression indicates that TVBS-II decay into both TVBS-I and TVFS, but cannot relax over the other TVBS-II. As a result, 
the rate for the decay of a given TVBS-II into a particular TVBS-I is approximately given by
\begin{equation}
W_{k\sigma=2\rightarrow k'\sigma=1} \approx\frac{W}{N}\mid \Psi_{k\sigma=2}(1)\mid^{2}\mid \Psi_{k'\sigma=1}(0)\mid^{2}
\label{eq:WTVBSII}
\end{equation}
At this step, by performing the sum over the wave vector $k'$ in Eq.(\ref{eq:WTVBSII}), the rate for the decay over all the TVBS-I is equal to $W$. Since the global relaxation rate for the TVBS-II is equal to $3W/2$, the rate for the relaxation over the TVFS continuum is equal to $W/2$. These results are in a rather good agreement with the numerical results shown in Fig. 5. However, when the nonlinearity slightly decreases, both the TVBS-I and TVBS-II wave functions delocalize so that $\Psi_{k\sigma=2}(1)$ and $\Psi_{k\sigma=1}(0)$ decrease. In that context, the relaxation rate for the decay of TVBS-II into TVBS-I decreases, as shown in Figs. 5 and 6, whereas the rate for the relaxation over TVFS increases. In the same way, the relaxation channel connected to the decay into the other TVBS-II opens because the wave function $\Psi_{k\sigma=2}(m)$ does not vanish any more for $m=0$ or $m=2$.

At this step, let us mention that the previous results address the fundamental question of the manifestation of the quantum localization, i.e. of the existence of quantum breathers.
Indeed, classical breathers describe time periodic and spatially localized nonlinear vibrations. By contrast, due to the translational invariance, the center of mass of the two vibrons in a TVBS is fully delocalized according to a Bloch wave. However, the localized nature of a TVBS arises due to the trapping of the two quanta around few sites. 
In that context, although a TVBS does not describe a spatially localized field, its degree of localization manifests itself through particular correlation functions \cite{kn:wang1,kn:wang2}. Our results clearly establish that such correlation functions are involved in the calculation of the relaxation rate (see for instance Eq. (\ref{eq:corr})). As a consequence, the dependence of the relaxation channels versus the nature of the two-vibron states can be viewed as a manifestation of the quantum localization of TVBS.
 
To conclude, let us note that the knowledge of the relaxation rates allow us to characterize the population dynamics of the two-vibron eigenstates in real time. To illustrate this feature, let us consider the strong nonlinear limit in which the relaxation schema is rather simple. Indeed, the rates are almost wave vector independent so that the system can be modeled as a three levels system formed by the TVBS-I band, the TVBS-II band and the TVFS band (see Fig. 7). The first relaxation channel corresponds to the decay of the TVBS-I band into the TVBS-II band according to the rate $W$. The relaxation of the TVBS-II band exhibits two main channels connected to the decay into the TVBS-I band with the rate $W$ and to the decay into the TVFS band with the rate $W/2$. Finally, TVFS are allowed to relax into the other TVFS, only. Let $P_{I}$, $P_{II}$ and $P_{F}$ denote the populations of these bands. Therefore, the time evolution of these populations is governed by a master equation expressed as 
\begin{eqnarray}
\dot{P_{I}}&=&-W P_{I}+W P_{II} \nonumber \\
\dot{P_{II}}&=&W P_{I}-\frac{3W}{2} P_{II} \nonumber \\
\dot{P_{F}}&=&\frac{W}{2} P_{II}
\label{eq:master}
\end{eqnarray} 
The system Eq.(\ref{eq:master}), which can be solved straightforwardly, allows us to follow the relaxation pathways in real time. For instance, if we assume that the system is initially set in the TVBS-I band, Eq.(\ref{eq:master}) clearly shows that the population $P_{I}(t)$ decays so that the population of the TVBS-II band increases. However, as $P_{II}(t)$ increases with time, it tends to relax into the TVFS continuum. Therefore, we observe the relaxation of the TVBS-I band into the TVFS continuum through the TVBS-II band. After a time $t^{*}=21.5/W$, both $P_{I}$ and $P_{II}$ can be neglected whereas $P_{F}$  is equal to $0.99$. At biological temperature ($T=310$ K), with the realistic parameters $A=8 $ cm$^{-1}$, $E_{B}=12$ cm$^{-1}$, and $J=8$ cm$^{-1}$ we obtain $W=21.36$ cm$^{-1}$ so that $t^{*}=5.3 $ps. 

Note that in a recent paper devoted to the TVBS relaxation in a molecular nanowire, it has been pointed out that, for a one-dimensional phonon bath, a given TVBS essentially decays into the other TVBS \cite{kn:pouthier2}. The origine of the different with the present work is twofold. First, in Ref.\cite{kn:pouthier2}, the vibron dynamics was described according to a Hubbard model for bosons which yields a single TVBS band (almost identical to the TVBS-I). Then, the coupling with the thermmal bath was assumed to be responsibel for a random modulation of the frequency of each molecule, i.e. $\Delta H \approx \sum \Delta \omega_{n} b_{n}^{+}b_{n}$, in marked contrast with the modulation of the vibron hopping constant considered in this paper (see Eq.(\ref{eq:DH})).

To summarize, the present paper was devoted to the characterization of the two-vibron relaxation mechanisms in an anharmonic $\alpha$-helix protein. According to the small polaron model developed in Ref. \cite{kn:pouthier4}, it was shown that the relaxation originates in the interaction between the dressed anharmonic vibrons and the remaining phonons. This interaction is responsible for the occurrence of transitions between two-vibron eigenstates mediated by both phonon absorption and phonon emission. 
At biological temperature, we have established that the relaxation rate does not depends significantly on the nature of the two-vibron state involved in the process. Therefore, for realistic values of the parameters, the rate ranges between 10 and 40 cm$^{-1}$ so that the corresponding lifetime is about 0.1 - 1.0 ps. Note that TVBS-II decay more rapidly than both TVBS-I and TVFS, the ratio between the corresponding rates being of about 1.5. 
By contrast, we have shown that the relaxation channels strongly depends on the nature of the two-vibron states. More precisely, the spatially localized nature of TVBS, i.e. their breather-like character, is responsible for the specification of a given relaxation pathway. 
In that context, TVBS-I, which correspond to the trapping of the two quanta around the same amide-I vibration, tend to decay into TVBS-II which refer to the trapping of the two quanta around two nearest neighbor amide-I vibrations. By contrast, TVBS-II decay into both TVBS-I and TVFS, the first channel being two times more efficient than the second one.

\begin{center}
\textbf{Figure Caption}
\end{center}

Figure 1 : Temperature dependence of the zero wave vector TVBS-I relaxation rate (full circles) for $A=8$ cm$^{-1}$ and for $E_{B}=4$ cm$^{-1}$ (a) and $E_{B}=12$ cm$^{-1}$ (b). Empty circles correspond to the rate for the relaxation over all the other TVBS-I whereas empty squares represent the rate for the decay into the set of the second eigenstates $\mid \Psi_{k\sigma=2} \rangle$ (see the text).

Figure 2 : TVBS-I relaxation rate (full circles) versus the intramolecular anharmonicity at $T=310$ K for for $E_{B}=4$ cm$^{-1}$ (a), $E_{B}=8$ cm$^{-1}$ (b) and
$E_{B}=12$ cm$^{-1}$ (c). Empty circles correspond to the rate for the relaxation over all the other TVBS-I whereas empty squares represent the rate for the decay into the set of the second eigenstates $\mid \Psi_{k\sigma=2} \rangle$ (see the text).

Figure 3 : TVBS-I relaxation rate (full circles) versus the small polaron binding energy at $T=310$ K for for $A=8$ cm$^{-1}$. Empty circles correspond to the rate for the relaxation over all the other TVBS-I whereas empty squares represent the rate for the decay into the set of the second eigenstates $\mid \Psi_{k\sigma=2} \rangle$ (see the text).

Figure 4 : Correlations between the relaxation channels and the nature of the two-vibron eigenstates for $T=310$ K, $A=8$ cm$^{-1}$ and $E_{B}=4$ cm$^{-1}$ (a) and $E_{B}=10$ cm$^{-1}$ (b). The upper panel represents two-vibron energy spectrum whereas the lower panel displays the wave vector dependence of the relaxation rates. Open circles characterize the rate $W_{0,1 \rightarrow k\sigma=1}$ whereas open squares correspond to the rate $W_{0,1 \rightarrow k\sigma=2}$. 

Figure 5 : Relaxation rate (full circles) of the $\mid \Psi_{k=0,\sigma=2} \rangle$ eigenstate versus $E_{B}$ at T=310 K and for $A=8$ cm$^{-1}$. Empty circles represent the rate for the decay into the TVBS-I and Empty squares correspond to the rate for the decay into all the other $\mid \Psi_{k,\sigma=2} \rangle$ eigenstates. Empty triangles characterize the rate for the decay into the TVFS continuum.

Figure 6 : Relaxation rate (full circles) of the $\mid \Psi_{k=0,\sigma=2} \rangle$ eigenstate versus $A$ at T=310 K and for $E_{B}=10$ cm$^{-1}$. Empty circles represent the rate for the decay into the TVBS-I and Empty squares correspond to the rate for the decay into all the other $\mid \Psi_{k,\sigma=2} \rangle$ eigenstates. Empty triangles characterize the rate for the decay into the TVFS continuum.

Figure 7 : Three levels model for the population dynamics of the two-vibron states within the strong nonlinear limit.

   \begin{figure}[p!]   
   \includegraphics{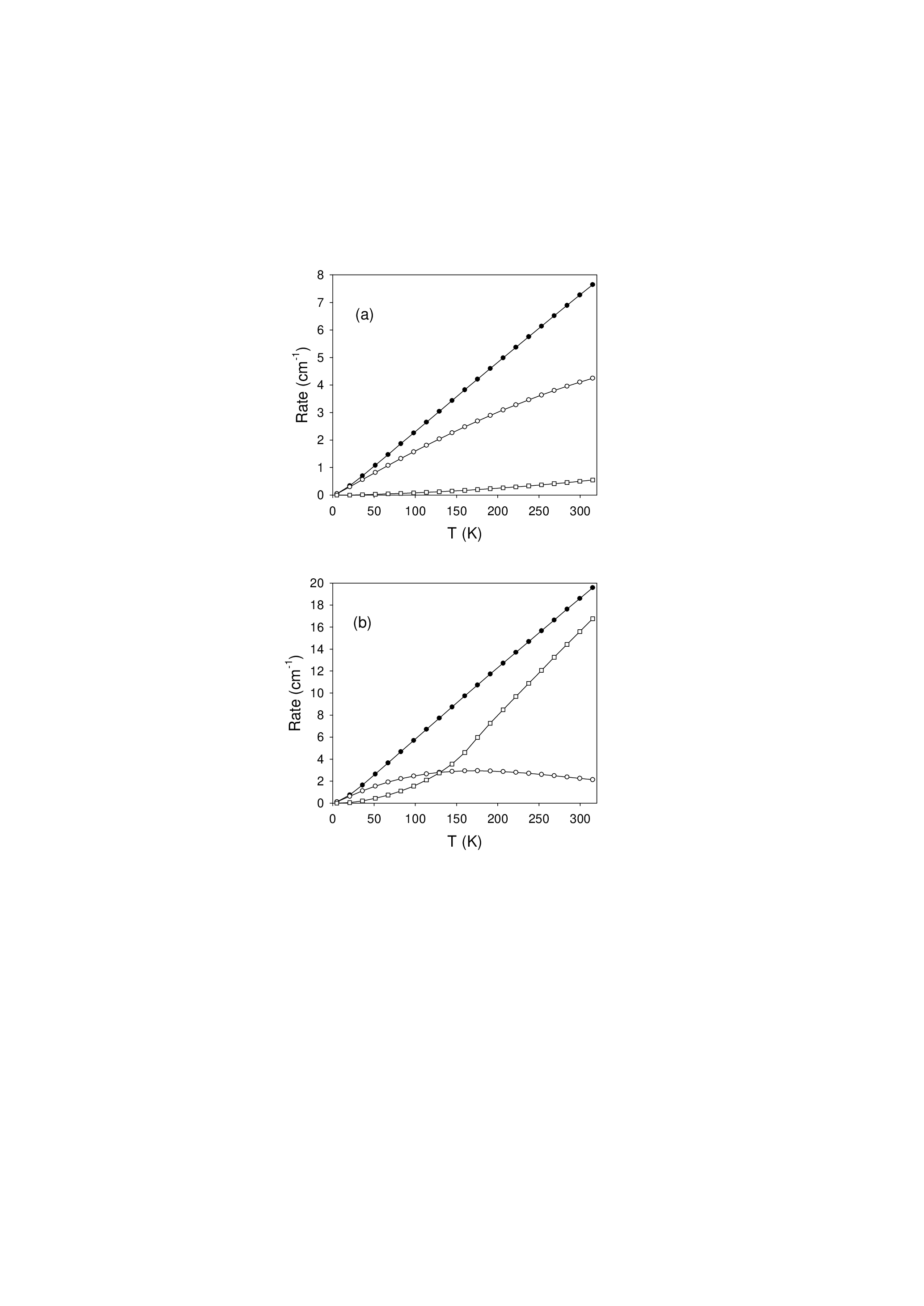}
   \caption[]{
  Temperature dependence of the zero wave vector TVBS-I relaxation rate (full circles) for $A=8$ cm$^{-1}$ 
and for $E_{B}=4$ cm$^{-1}$ (a) and $E_{B}=12$ cm$^{-1}$ (b). Empty circles correspond to the rate for the 
relaxation over all the other TVBS-I whereas empty squares represent the rate for the decay into the set of the second 
eigenstates $\mid \Psi_{k\sigma=2} \rangle$ (see the text).
   }\label{fig:figure1}
   \end{figure}

\begin{figure}[p!]   
\includegraphics{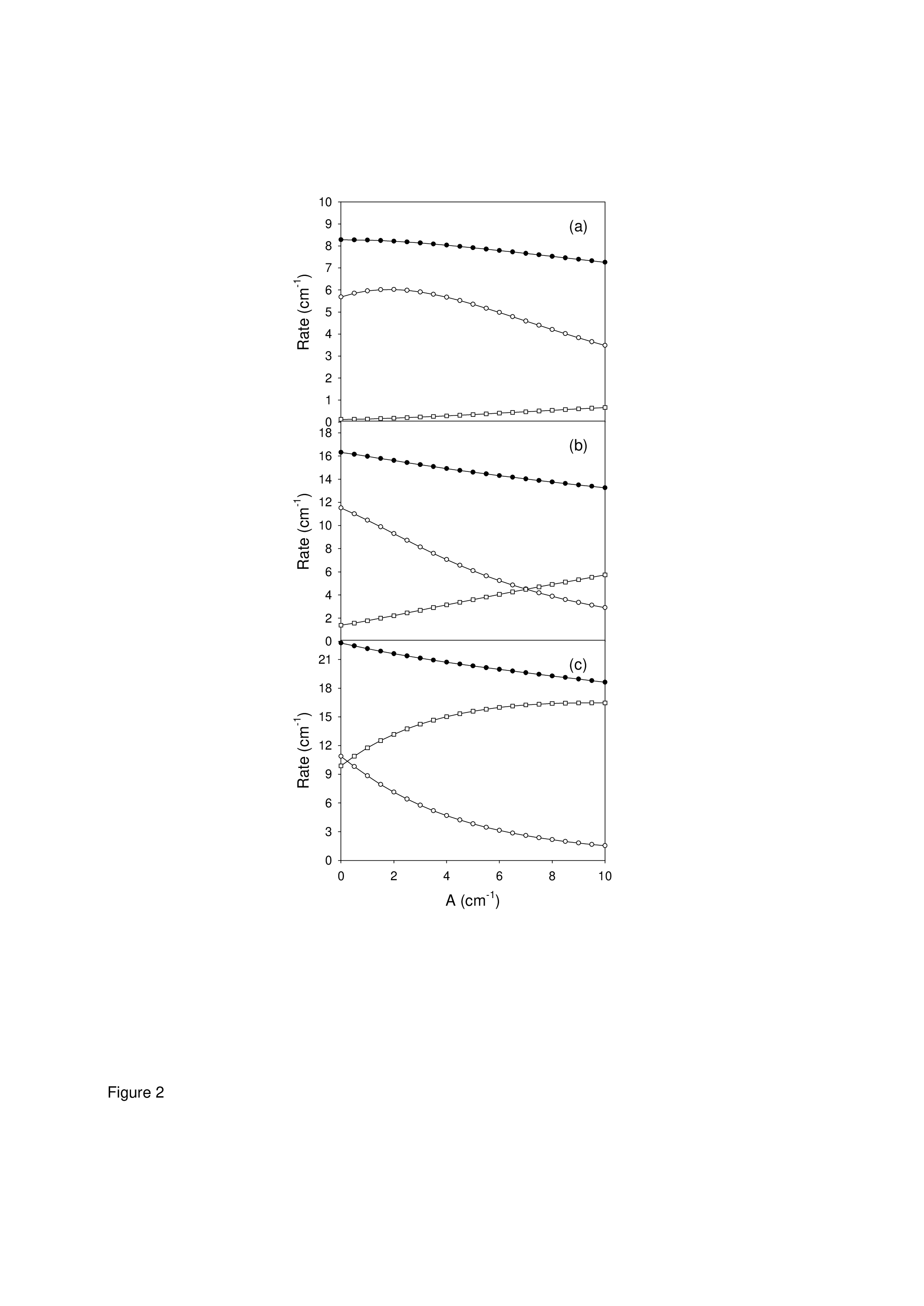}
 \caption[]{TVBS-I relaxation rate (full circles) versus the intramolecular anharmonicity 
at $T=310$ K for for $E_{B}=4$ cm$^{-1}$ (a), $E_{B}=8$ cm$^{-1}$ (b) and $E_{B}=12$ cm$^{-1}$ (c). 
Empty circles correspond to the rate for the relaxation over all the other TVBS-I whereas empty squares 
represent the rate for the decay into the set of the second eigenstates $\mid \Psi_{k\sigma=2} \rangle$ (see the text).
}\label{fig:figure2}
\end{figure}

\begin{figure}[p!]   
\includegraphics{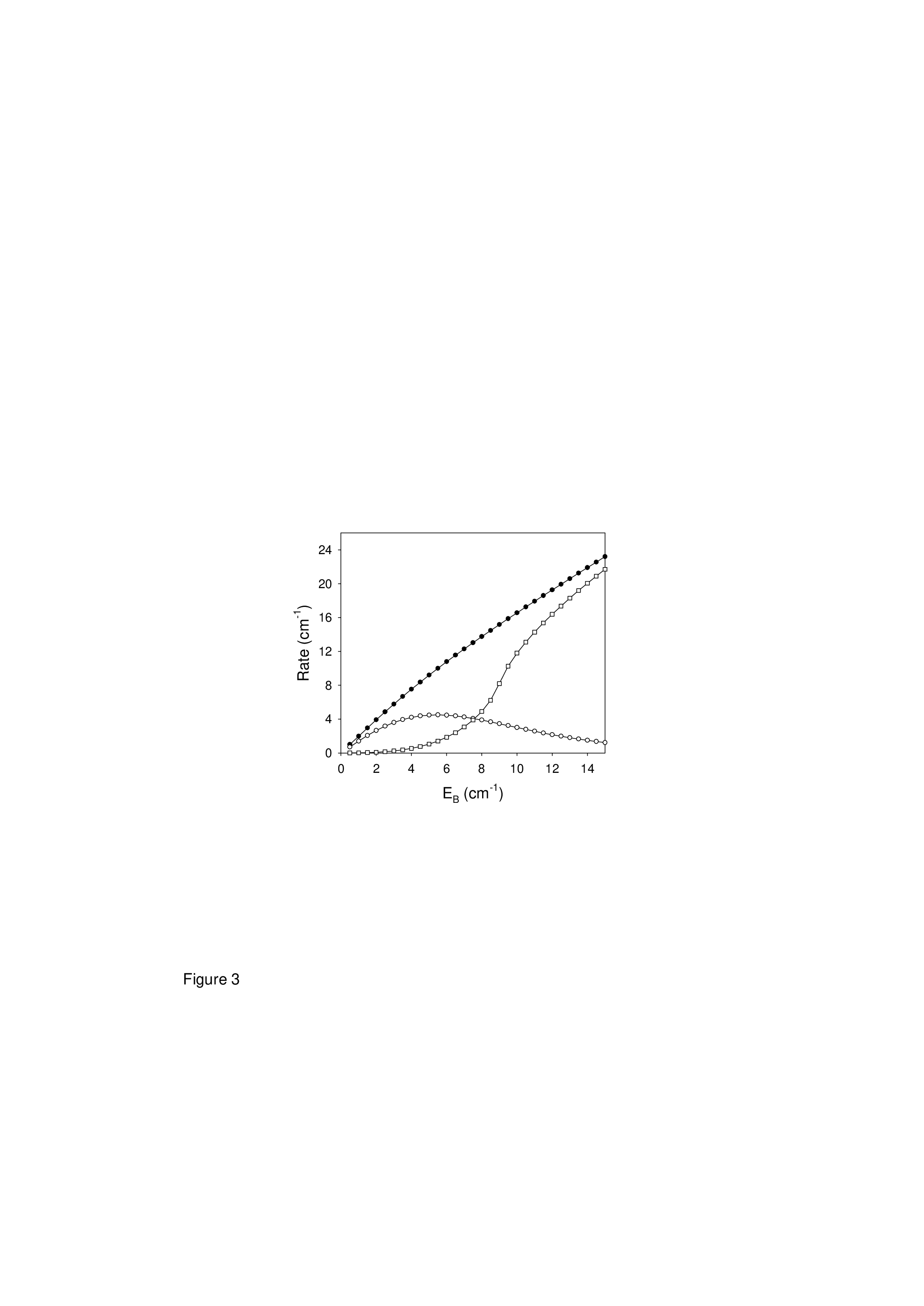}
 \caption[]{TVBS-I relaxation rate (full circles) versus the small polaron binding energy 
at $T=310$ K for for $A=8$ cm$^{-1}$. Empty circles correspond to the rate for the relaxation 
over all the other TVBS-I whereas empty squares represent the rate for the decay into the set of the 
second eigenstates $\mid \Psi_{k\sigma=2} \rangle$ (see the text).
}\label{fig:figure3}
\end{figure}

\begin{figure}[p!]   
\includegraphics{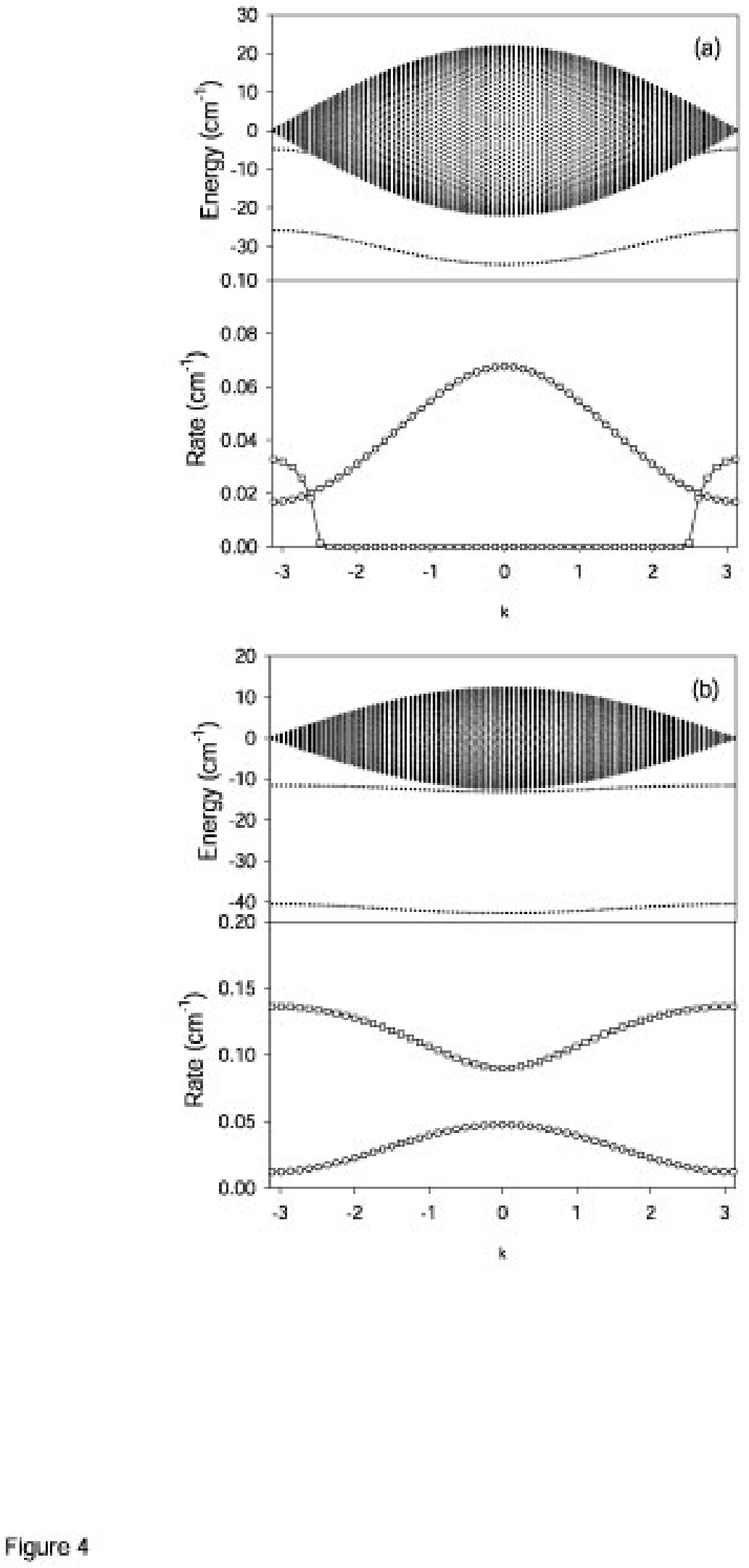}
 \caption[]{Correlations between the relaxation channels and the nature of the two-vibron eigenstates 
for $T=310$ K, $A=8$ cm$^{-1}$ and $E_{B}=4$ cm$^{-1}$ (a) and $E_{B}=10$ cm$^{-1}$ (b). 
The upper panel represents two-vibron energy spectrum whereas the lower panel displays the wave 
vector dependence of the relaxation rates. Open circles characterize the rate $W_{0,1 \rightarrow k\sigma=1}$ 
whereas open squares correspond to the rate $W_{0,1 \rightarrow k\sigma=2}$.
}\label{fig:figure4}
\end{figure}

\begin{figure}[p!]   
\includegraphics{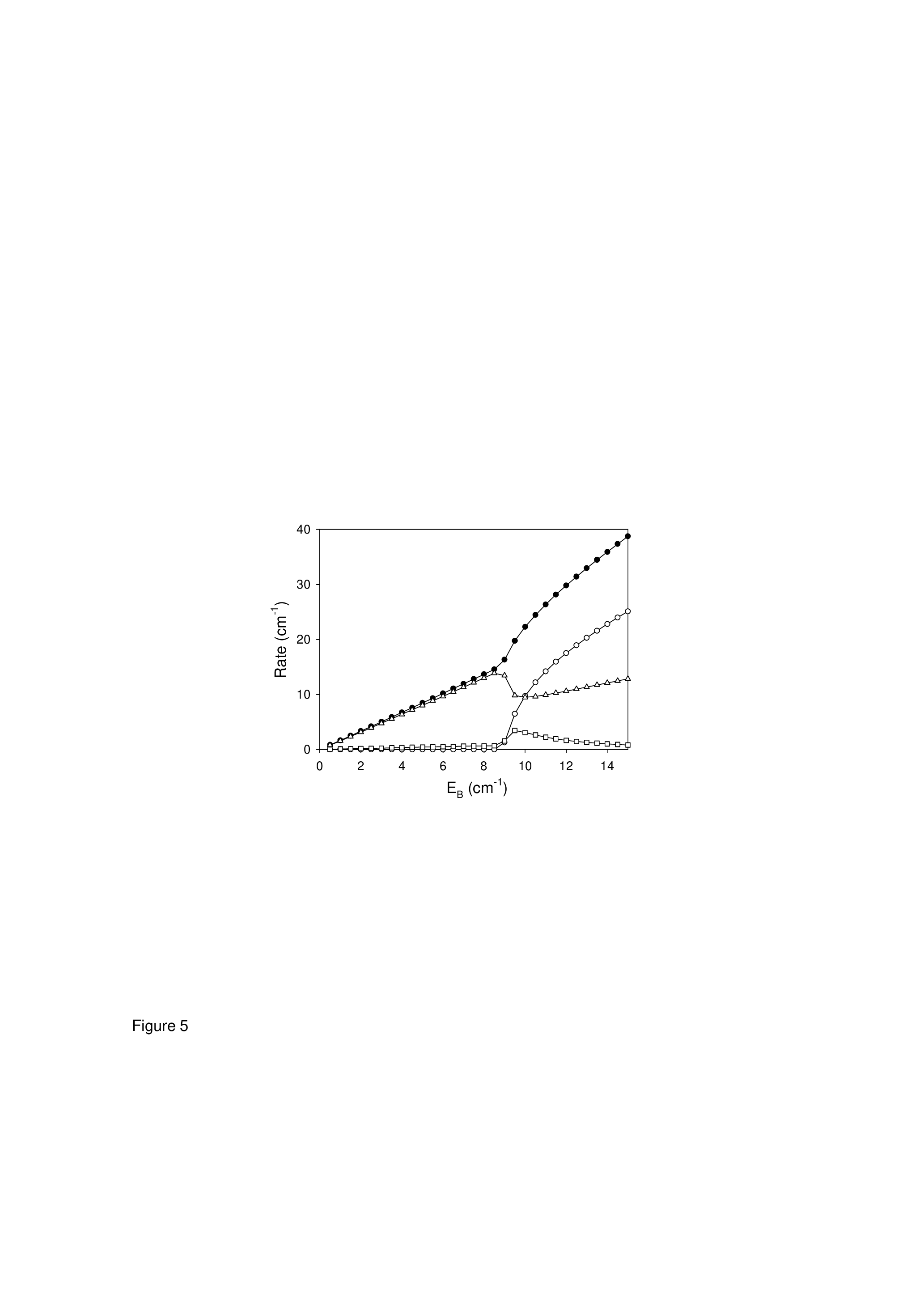}
 \caption[]{Relaxation rate (full circles) of the $\mid \Psi_{k=0,\sigma=2} \rangle$ eigenstate 
versus $E_{B}$ at T=310 K and for $A=8$ cm$^{-1}$. Empty circles represent the rate for 
the decay into the TVBS-I and Empty squares correspond to the rate for the decay into all 
the other $\mid \Psi_{k,\sigma=2} \rangle$ eigenstates. Empty triangles characterize the rate
for the decay into the TVFS continuum.
}\label{fig:figure5}
\end{figure}

\begin{figure}[p!]   
\includegraphics{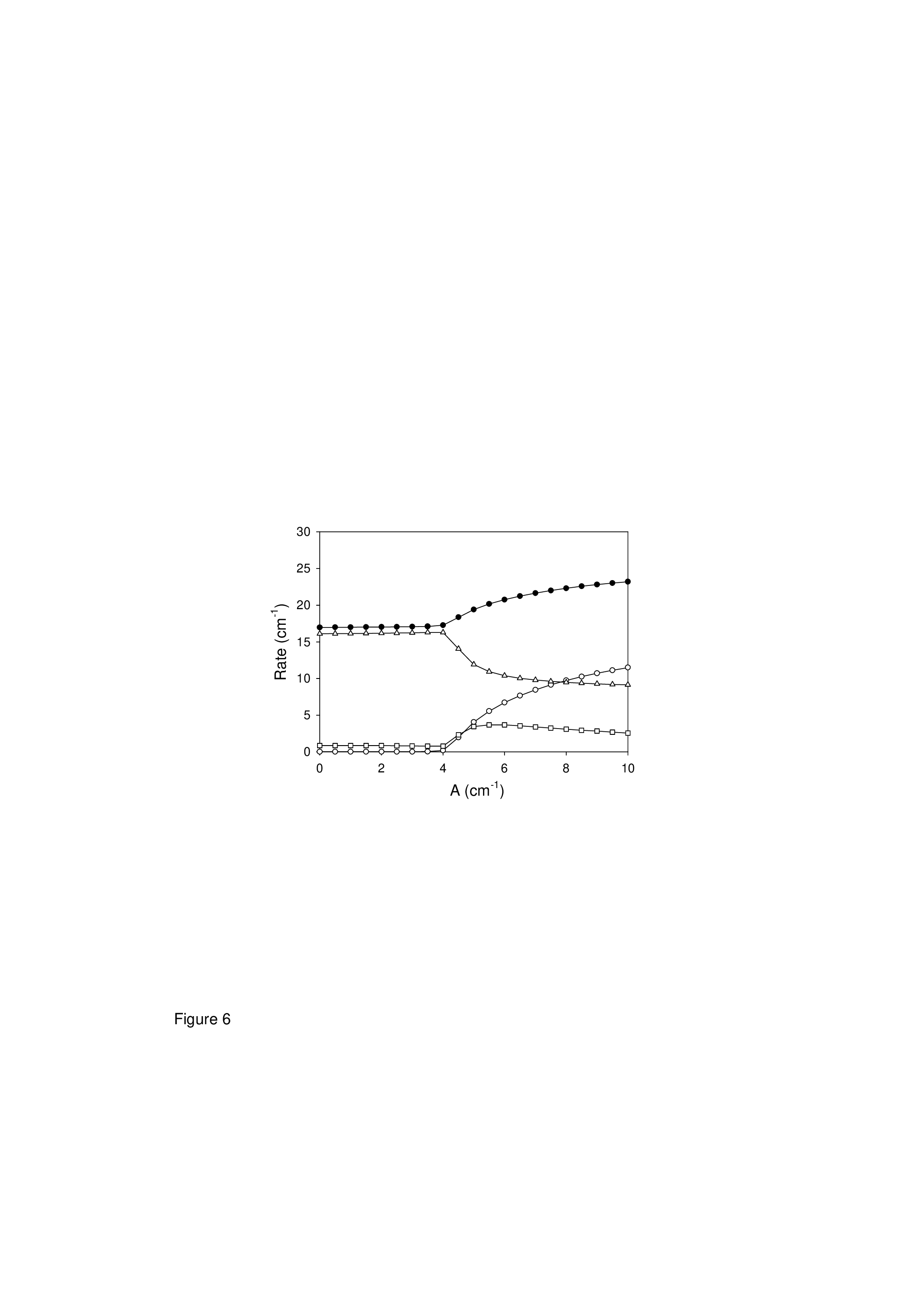}
 \caption[]{Relaxation rate (full circles) of the $\mid \Psi_{k=0,\sigma=2} \rangle$ eigenstate
versus $A$ at T=310 K and for $E_{B}=10$ cm$^{-1}$. Empty circles represent the rate 
for the decay into the TVBS-I and Empty squares correspond to the rate for the decay into 
all the other $\mid \Psi_{k,\sigma=2} \rangle$ eigenstates. Empty triangles characterize the 
rate for the decay into the TVFS continuum.
}\label{fig:figure6}
\end{figure}

\begin{figure}[p!]   
\includegraphics{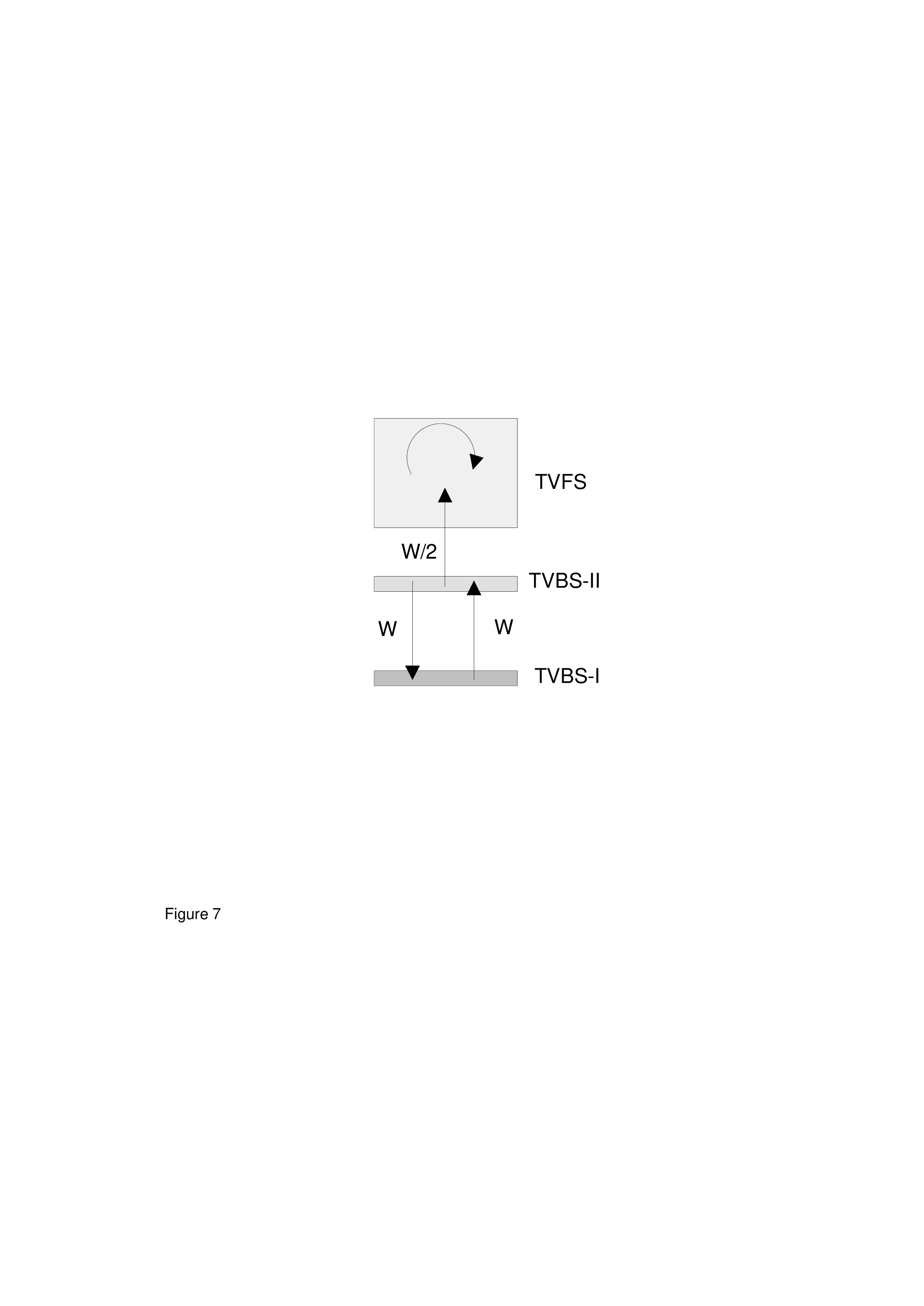}
 \caption[]{Three levels model for the population dynamics of the two-vibron states within the 
strong nonlinear limit.
}\label{fig:figure7}
\end{figure}


\begin{thebibliography}{99}
\bibitem{kn:davydov}  A. S. Davydov and N. I. Kisluka, Phys. Status Solidi \textbf{59}, 465 (1973); Zh. Eksp. Teor. Fiz \textbf{71}, 1090 (1976) [Sov. Phys. JETP \textbf{44}, 571 (1976)].
\bibitem{kn:scott1} A.C. Scott, Phys. Rep. \textbf{217}, 1 (1992).
\bibitem{kn:chris} P. L. Christiansen and A. C. Scott , \emph{Davydov's Soliton Revisited},(Plenum, New York, 1990).
\bibitem{kn:brown1}D.W. Brown and Z. Ivic, Phys. Rev. \textbf{B40}, 9876 (1989). 
\bibitem{kn:brown2}D.W. Brown, K. Lindenberg, and X. Wang, in \emph{Davydov's Soliton Revisited},(Plenum, New York, 1990), edited by P. L. Christiansen and A. C. Scott.
\bibitem{kn:ivic1}Z. Ivic, D. Kapor, M. Skrinjar, and Z. Popovic, Phys. Rev. \textbf{B48}, 3721 (1993).
\bibitem{kn:ivic2}Z. Ivic, D. Kostic, Z. Przulj, and D. Kapor, J. Phys. Condens. Matter \textbf{9}, 413 (1997).
\bibitem{kn:ivic3}J. Tekic, Z. Ivic, S. Zekovic, and Z. Przulj, Phys. Rev. \textbf{E60}, 821 (1999).
\bibitem{kn:ivic4}Z. Ivic, Z. Przulj, and D. Kostic, Phys. Rev. \textbf{E61}, 6963 (2000).
\bibitem{kn:kimball} J. C. Kimball, C. Y. Fong, and Y. R. Shen, Phys. Rev. \textbf{B23}, 4946 (1981).
\bibitem{kn:bogani} F. Bogani, G. Cardini, V. Schettino, and P. L. Tasselli ,  Phys. Rev. \textbf{B42}, 2307 (1990).
\bibitem{kn:scott} A. C. Scott, J. C. Eilbeck, and H. Gilhoj, Physica \textbf{D78}, 194 (1994).
\bibitem{kn:pouthier1} V. Pouthier and C. girardet, Phys. Rev. \textbf{B65}, 035414 (2002).
\bibitem{kn:pouthier2} V. Pouthier, J. Chem. Phys. \textbf{118}, 3736 (2003).
\bibitem{kn:pouthier3} V. Pouthier, J. Chem. Phys. \textbf{118}, 9364 (2003).
\bibitem{kn:pouthier4} V. Pouthier, Phys. Rev. \textbf{E68}, 021909 (2003).
\bibitem{kn:aubry} S. Aubry, Physica \textbf{D103}, 201 (1997).
\bibitem{kn:flach} S. Flach and C.R. Willis, Phys. Rep. \textbf{295}, 181 (1998). 
\bibitem{kn:mackay} R.S. MacKay, Physica \textbf{A288}, 174 (2000).
\bibitem{kn:guyot1} P. Guyot-Sionnest, Phys. Rev. Lett. \textbf{67}, 2323 (1991).
\bibitem{kn:sih} R. Honke, P. Jakob, Y. J. Chabal, A. Dvorak, S.Tausendpfund, W. Stigler, P. Pavone, A. P. Mayer, and U. Schr$\ddot{o}$der, Phys. Rev. \textbf{B59}, 10996 (1999). 
\bibitem{kn:shen} R. P. Chin, X. Blase, Y. R. Shen, and S. GT. Louie, Europhys. Lett. \textbf{30}, 399 (1995).
\bibitem{kn:jakob1} P. Jakob, Phys. Rev. Lett. \textbf{77}, 4229 (1996).
\bibitem{kn:jakob2} P. Jakob, Physica \textbf{D119}, 109 (1998).
\bibitem{kn:jakob3} P. Jakob, J. Chem. Phys. \textbf{114}, 3692 (2001).
\bibitem{kn:jakob4} P. Jakob, Appl. Phys. \textbf{A75}, 45 (2002).
\bibitem{kn:jakob_p1} P. Jakob and B.N.J. Persson, J. Chem. Phys. \textbf{109}, 8641 (1998).
\bibitem{kn:okuyama} H. Okuyama, T. Ueda, T. Aruga, and M. Nishijima, Phys. Rev. \textbf{B63}, 233404 (2001).
\bibitem{kn:cruzeiro1} L. Cruzeiro-Hansson, Phys. Lett. A 249 465-473 (1998); 
Phys. Lett. A 223 383-388 (1996). 
\bibitem{kn:cruzeiro2} L. Cruzeiro-Hansson and S. Takeno, Phys. Rev. \textbf{E56}, 894 (1997). 
\bibitem{kn:cruzeiro3} L. Cruzeiro-Hansson, Phys. Rev. Lett. \textbf{73}, 2927 (1994). 
\bibitem{kn:forner} W. Forner, J. Phys. Condens. Matter \textbf{3}, 3235 (1991). 

\bibitem{kn:cottingham} J. P. Cottingham and J. W. Schweitzer, Phys. Rev. Lett. \textbf{62}, 1792 (1989).
\bibitem{kn:schweitzer} J. W. Schweitzer, Phys. Rev. \textbf{A45}, 8914 (1992).
\bibitem{kn:hamm1} P. Hamm, M. Lim, and R. M. Hochstrasser J. Phys. Chem. \textbf{B102}, 6123 (1998).
\bibitem{kn:hamm2} S. Woutersen and P. Hamm, J. Phys.: Condens. Matter \textbf{14}, R1035 (2002).
\bibitem{kn:wang1} W. Z. Wang, J. T. Gammel, and A. R. Bishop, Phys. Rev. Lett. \textbf{76}, 3598 (1996).
\bibitem{kn:wang2} W. Z. Wang, A. R. Bishop, J. T. Gammel, and R. N. Silver, Phys. Rev. Lett. \textbf{80}, 3284 (1998).
\end{thebibliography}
\end{document}